\documentstyle{neu}
\begin{document}

\title{NONTHERMAL RADIATION OF THE CRAB NEBULA}

\author{F.A. Aharonian  \\
{\it    Max Planck Institut f\"ur Kernphysik,   
Postfach 103980, \\
D-69029 Heidelberg, Germany, aharon@fel.mpi-hd.mpg.de}\\
A.M. Atoyan\\
{\it Yerevan Physics Institute, Armenia, atoyan@boris.mpi-hd.mpg.de}}

\maketitle

\small 

\section*{Abstract}

The radiation mechanisms contributing to formation 
of the nonthermal spectrum of the Crab Nebula, 
as well as the information that could be derived from
future observations in different energy bands, 
are  discussed.

\section{Introduction}

The Crab Nebula is a unique cosmic laboratory 
with an unprecedentedly broad spectrum of the observed
nonthermal radiation which extends throughout  21 (!) decades of 
frequences -- from radio wavelengths  to very high energy $\gamma$-rays 
(see Fig.~1). This emission  is dominated by 
two major mechanisms connected with interactions of 
relativistic electrons  with the nebular magnetic and photon fields.   
While the synchrotron component  is responsible for 
the radiation from radio to relatively low energy 
$\gamma$-rays ($E < 1 \, \rm GeV$), the inverse Compton (IC) scattering
of electrons is thought as the most probable  
mechanism for very high energy  $\gamma$-rays discovered 
in the late 80's by the Whipple collaboration (Weekes et al. 1989).  

\begin{figure}[t]
\vspace{5.4 cm}
\includegraphics{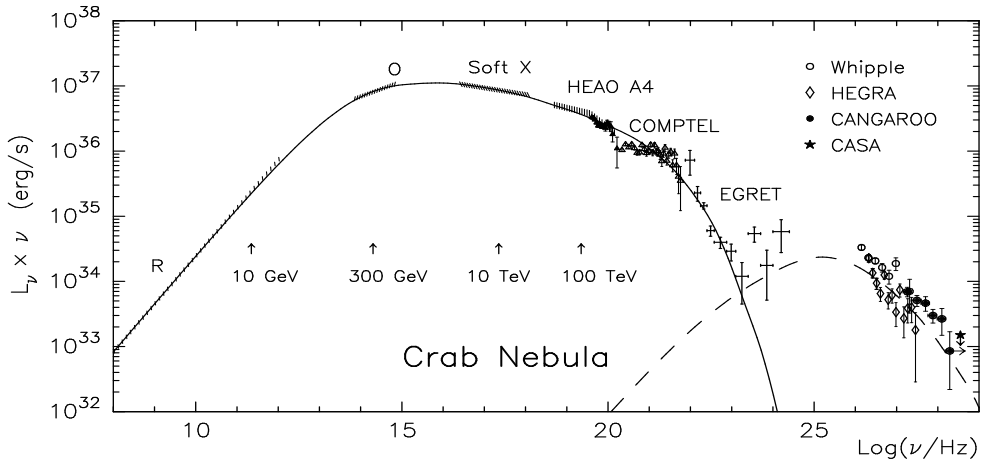}
\caption{\protect \footnotesize Nonthermal radiation of the  
Crab Nebula. In the $\gamma$-ray domain
only the recently reported fluxes are shown: 
COMPTEL (van der Meulen et al. 1998), EGRET (Fierro 1996),
Whipple (Mohanty et al. 1998), HEGRA (Aharonian et al. 1997),
CANGAROO (Tanimori et al. 1998), CASA (Borione et al. 1997).  
The solid and dashed curves correspond 
to the synchrotron and inverse Compton components of radiation, respectively, 
calculated in the framework of the spherically symmetric 
MHD wind model of KC84.}
\end{figure}

The typical energies of electrons responsible for production of 
synchrotron photons in different energy bands of the Crab spectrum  
are indicated in Fig.~1.  Note that although the conclusion about 
highest energy electrons is still based on a model (although well justified) 
assumption about the  synchrotron  origin of the 
hard X-rays/low-energy $\gamma$-rays, 
the detection of $\gamma$-rays well above $10^{13} \, \rm eV$ is an 
{\it unambiguous} evidence  of effective acceleration of particles 
beyond $10^{14} \, \rm eV$. 

It is commonly accepted that the synchrotron nebula is powered by 
the relativistic wind of electrons generated at the pulsar and
terminated by a standing reverse shock wave at a distance 
$r_{\rm s} \sim 0.1 \, \rm pc$ (Rees and  Gun 1974).    
The relativistic MHD models,  
even in their simplified form (e.g. ignoring the 
axisymmetric structure  of the wind  and its interaction with the optical 
filaments),  successfully describe 
the general characteristics of the synchrotron nebula, and
predict  realistic distributions of relativistic
electrons and magnetic field 
in the downstream region behind the shock (e.g. Kennel and Coroniti 1984,
hereafter KC84).

While  the synchrotron and IC mechanisms seem to provide a reasonable
explanation of the overall nonthermal radiation of the  
Crab Nebula (see Fig.~1), one cannot exclude possible deviations 
in different frequency  domains  from the  
simplified picture of the outer  
nebula described by the  spherically symmetric MHD models. 
Indeed, the  
recent imaging of the Crab Nebula by the Hubble Space Telescope and ROSAT  
revealed a very complex structure of  the inner part of the nebula 
consisting of features like wisps,  jets, knots, {\it etc.} 
(Hester et al. 1995). 
This shows that we are far from complete  
understanding of the physics of interaction of the pulsar wind with the 
synchrotron nebula, and
ensures that this object remains a great challenge for future  
theoretical work. 
Although it seems unlikely that the incorporation of the hot spots would 
dramatically change the overall nonthermal spectrum of the Crab Nebula,
nevertheless they could introduce non-negligible spectral features
in the X-ray and $\gamma$-ray domains.
Furthermore, a significant production of high-energy 
$\gamma$-rays, on top of the IC radiation originating in the  
synchrotron nebula,  could occur in the filaments, provided that 
the relativistic particles were partially confined in these high 
density regions. Within this scenario, both the bremsstrahlung
of relativistic electrons, and $\pi^0$-decay $\gamma$-rays 
produced by  protons
might significantly affect the overall $\gamma$-ray spectrum.

\section{Synchrotron and IC radiation of the Crab Nebula}
      
To calculate the nonthermal radiation of the Crab Nebula, 
one has to specify the 
spatial distribution of the magnetic field, the 
acceleration site(s) and the injection spectrum of
the relativistic electrons, as well as  the character of their propagation 
in the nebula.   
Remarkably, the MHD model of KC84 provides all these parameters, in particular
allows  self-consistent  calculation of  the
spatial and spectral distribution of high energy electrons, freshly 
accelerated at the wind termination shock and injected 
into the nebula ({\it wind} electrons).  Arons and co-workers 
(see e.g. Arons 1996) have shown that the wind termination shock 
is able to accelerate particles  up to $\sim 10^{15} \, \rm eV$.

Meanwhile the radio emission of the 
nebula  requires an additional  low energy ($E \leq 200$~GeV) component
of electrons ({\it radio} electrons) accumulated, most probably, 
during the whole history of the Crab (KC84).
Although the origin and site(s) of  this ``relic'' component 
are not yet established, it can be easily incorporated into
the calculations of the broad-band 
synchrotron spectrum with a minimum number 
of assumptions based on radio observations (Atoyan and Aharonian 1996, 
hereafter AA96).

Fig.~1 demonstrates good  agreement of calculations 
with the observed spectrum of the Crab
Nebula up to hard X-rays, and a reasonable explanation of the 
$\gamma$-ray fluxes  up to 1 GeV by the synchrotron radiation. 
The best fit is reached by the following 
combination of the  spectra of the {\it radio} 
and {\it wind}  electrons:
(1) $n_{\rm re}(E) \propto E^{-\alpha_{\rm re}} \exp(-E/E_{\rm c})$ with
$\alpha_{\rm re}= 1.52$ and $E_{\rm c} = 150 \, \rm GeV$, and 
(2) $n_{\rm we}(E) \propto (E_\ast+E)^{-\alpha_{\rm we}} \exp(-E/E_{\rm 1})$ 
with $\alpha_{\rm we}=2.4$,
$E_\ast=200 \, \rm GeV$,  and $E_1 = 2.5 \cdot 10^{15}  \, \rm eV$. 
The presentation of the spectrum of the {\it wind} electrons 
$n_{\rm we}(E)$ 
in such form  provides a necessary (and natural  
in the framework of MHD model) 
flattening of the spectrum below $E_\ast \sim 200 \, \rm GeV$. 
The transition from the hard to steep power-laws in the total 
({\it radio} +
{\it wind}) electron spectrum at energies around 
100~GeV accounts for the sharp
steepening of the spectrum at IR/optical
wavelengths.  Therefore detailed spectroscopic measurements in this 
energy region would allow us to specify more precisely the
values of $E_{\rm c}$ and $E_\ast$ which define the degree of ``smoothness'' 
of transition from {\it radio}  to {\it wind} electrons. An
independent information about the electrons
in this transition region is contained in the 
10 to 100 GeV $\gamma$-rays produced by the same electrons upscattering
the ambient low-frequency radiation.
Meanwhile, determination of the high energy 
cutoff $E_{\rm 1}$ in $n_{\rm we}(E)$ 
is contingent of spectroscopic measurements of $\gamma$-rays in the 
1 MeV to 1 GeV energy band. 
                                  
The existence of ultra-relativistic electrons
in the synchrotron nebula provides production
of detectable TeV $\gamma$-ray fluxes through the IC scattering (Gould 1965).
Since the target radiation fields which play major role in
the production of IC $\gamma$-rays of the Crab Nebula
(synchrotron, thermal far IR, and 2.7 K
background radiations) are well known,  the flux of
IC $\gamma$-rays can be calculated with good accuracy.
For the given flux of synchrotron X-rays, 
the number of TeV electrons strongly depends 
on the nebular magnetic field. Therefore the 
IC $\gamma$-ray fluxes are very sensitive to the average magnetic field:
$I(\geq 1 \, \rm TeV) \simeq 8 \cdot 10^{-12}
(\bar{B}/0.3 \, mG)^{-2.1} \, \rm ph/cm^2 s$. Thus the comparison of
the predicted and
observed TeV $\gamma$-ray fluxes allows determination of the magnetic field
in the central $r \sim 0.5 \, \rm pc$ region
(where the bulk of X-rays and TeV $\gamma$-rays are produced)
with accuracy $\Delta B /\bar{B} \simeq 0.5 (\Delta I /I)$.
Although the statistical significance of $\gamma$-ray 
observations of the Crab Nebula
is very high, the systematic uncertainties in the
flux estimates  remain rather large, $\Delta I/I \sim 1$ (see Fig.1).
Even with present uncertainties, the TeV observations favor the
magnetic field in the X-ray production region  
within $\bar{B} \sim 0.2-0.3 \,\rm mG$, which is in agreement with the
estimated equipartition field (Marsden et al. 1984).    

The shape of the spectrum of IC $\gamma$-rays is
rather stable to the basic parameters of the nebula  (AA96), in
particular, it is almost independent 
of the parameter $\sigma$ (the ratio of the electromagnetic 
energy flux to the particle energy flux  at the shock) which in
the framework of MHD model defines the spatial distribution 
of the magnetic field  $B(r)$ at $r \geq r_{\rm s}$.
Although the calculations in Fig.~1 correspond to  
$\sigma=0.005$,  the measured synchrotron fluxes can be equally well 
fitted by the $\sigma$ between 0.001 and 0.01. 
Indeed, since at distances  $\geq 3 r_{\rm s}$, where the 
bulk of synchrotron and IC photons are produced,
the magnetic field $B(r)$ depends rather weakly on $\sigma$ (see KC84),
only minor changes in the assumed injection spectrum of electrons 
are required to provide the same synchrotron spectrum for different 
$\sigma$ within 0.001-0.01.  Consequently, as 
far as the target photon fields for IC $\gamma$-rays are 
fixed, there is no room for strong dependence of the calculated 
IC fluxes on $\sigma$. 

This conclusion opposes to the statement of  De Jager and Harding (1992) 
and De Jager et al (1996; hereafter DJ96) about significant dependence of 
IC $\gamma$-ray  fluxes on $\sigma$. The method of calculations
of IC radiation used in these papers is based on the 
extraction of the ``prompt'' spectrum of electrons  in the nebula,
$N_{\rm e}(E, r)$, using
the spatial and energy distribution of the synchrotron 
radiation $F(\nu, r)$,  with an additional assumption about the 
structure of magnetic field $B(r)$ taken from KC84. Although 
quite appropriate  in principle, this method seems to be
not well justified, at least presently,  taken the fact
of the lack of proper information on $F(\nu, r)$, especially 
at X-rays -- the counterparts of TeV $\gamma$-rays. 
Such uncertainties leave large room for ambiguities 
in the $N_{\rm e}(E,r)$, and 
consequently in the IC $\gamma$-ray spectra. Therefore
it is crucial to control  
the accuracy of the procedure by comparison of the calculated 
synchrotron spectra (based on the  
{\it approximately} reconstructed $N_{\rm e}(E,r)$)  with 
the observed ones. Since 
such comparison is missing in the above mentioned papers,
we can only guess that the claimed
strong $\sigma$-dependence of the TeV fluxes does not exceed the
accuracy of calculations.
In this context, it is  worth noticing
that in DJ96, where the authors suggested an interesting idea 
of the synchrotron radiation due to 
the {\it second} population of highest 
energy electrons (see below), they show (Figure~1 in DJ96)
the IC spectra calculated for different $\sigma$  without 
presentation of the relevant synchrotron spectra
of the {\it main} electron component. On the other hand  
in the same Figure~1 of DJ96 is shown the hard 
synchrotron spectrum (``MeV bump'') produced by the second 
population of electrons, 
but without indication of contribution of the latter 
to the  IC $\gamma$-ray fluxes.

In Fig.~1 we show the combined synchrotron and IC components of 
radiation, calculated  in the framework 
of spherically symmetric  MHD model, which describe fairly 
well the flux levels of the Crab 
Nebula over the whole range of observed frequences. Meanwhile  
the ``zoom'' of Fig.~1 reveals some peculiarities in the spectrum,
in particular at 1-10 MeV, 1-10 GeV, and perhaps also at 
$\geq 10$~ TeV energies which could not be easily interpreted within the 
simplified synchrotron-Compton model. Below we discuss possible ways 
to account for these features.

\section{Second High Energy Synchrotron Component}

The recent spectral measurements of  the unpulsed radiation of the Crab
by COMPTEL revealed an unexpected flattening of the spectrum at  energies 
1-10~MeV (van der Meulen et al. 1998) 
which follows after well established  steepening of the spectrum
above 100 keV (Jung  1989, Bartlett et al. 1996). 
Since such sharp feature hardly could be attributed to    
peculiarities in the injection spectrum of shock accelerated electrons,
more natural interpretation of this spectral feature could be 
given assuming  the existence of 
an additional radiation component. Explanation of this
radiation excess in terms of nuclear $\gamma$-ray line emission is
not supported by observations  (van der Meulen et al. 1998), 
and more importantly, it contradicts to the total luminosity of the Crab
Nebula since only $\xi \leq 10^{-4}$ part of the energy losses of 
nonrelativistic protons and nuclei is released in prompt 
$\gamma$-ray lines, while its  main part goes to heating of the 
ambient gas, and thus should be shown up in the form of thermal radiation 
with unacceptably high luminosity   
$L_{\rm thermal} \sim \xi^{-1} L_{\rm obs} 
(1-10 \, {\rm MeV}) \geq 10^{40} \, \rm erg/s$.

\begin{figure}[t]
\vspace{5.5 cm}
\includegraphics{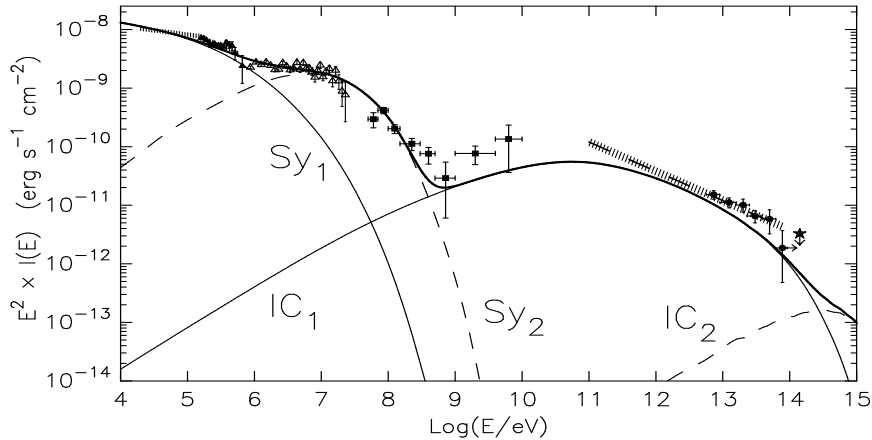}
\caption{\protect \footnotesize Synchrotron and IC radiation components   
produced by the first (solid) and second (dashed) 
populations of electrons (see text).
The heavy solid line shows the total flux. The hatched region corresponds to 
$I(E)=(2.5 \pm 0.4) \,   (E/1 \, \rm TeV)^{-2.5}
\, \rm cm^{-2} \, s^{-1} TeV^{-1}$ which generally describes the flux levels
from  300 GeV to 70 TeV reported by different groups.}
\end{figure}

While remaining in the framework of the hypothesis of the synchrotron origin 
of the radiation up to 1 GeV, the steepening 
above 100 keV implies an 
exponential cutoff  in the injection spectrum of the
{\it wind} electrons  at energies smaller than  
$E_1 = 2.5 \cdot 10^{15} \, \rm eV$
used in Fig.~1. If so,  the flat spectrum observed by COMPTEL
requires  the second population of high energy electrons (DJ96). 
In Fig.~2 we present a possible fit of the observed fluxes up to
1~GeV by two-component synchrotron emission. The first component
is attributed to the same {\it wind} electrons as in Fig.~1 but with 
$E_1 = 5 \cdot 10^{14} \, \rm eV$. For the second component we have to 
suppose very hard acceleration spectrum, for example of Maxwellian type,   
$n_2(E) \propto E^2 \exp(-E/E_2)$ with $E_2=5.5 \cdot 10^{14} \, \rm eV$.
Although  here $E_1 \approx E_2$,
actually  the mean energy of the second population 
$\bar{E}=3 E_2 \simeq 1.6 \cdot 10^{15} \, \rm eV$ is 
larger that the highest energy particles $E \sim E_1$ 
effectively present in the first population.  
Note that the needed acceleration power in the second electron population 
is only $P_2 \sim L_{\rm obs}(1-10 \, {\rm MeV}) \sim 10^{36} \, \rm erg/s$,
i.e. less than $1 \%$ of the first ({\it main}) population, 
$P_1 \sim 3 \cdot 10^{38} \, \rm erg/s$. 
This explains the small contribution of the second 
electron population into the total IC $\gamma$-ray fluxes up to energies 
$10^{14} \, \rm eV$ (see Fig.~2).

The possible sites of acceleration of the second electron 
population  could be the peculiar compact regions 
such as wisps, knots, etc. Since the equipartition magnetic field
in these regions is estimated  as high as few mG
(in that case $E_2$ is  reduced to $\simeq 10^{14} \, \rm eV$)  
the highest energy electrons could not escape the
acceleration sites due to severe synchrotron losses.
Although these
highly variable structures  
with typical size $0.2^{\prime \prime}$ (Hester 1995) 
are not resolvable by low-energy $\gamma$-ray detectors,
the confirmation of variability of 1-150~MeV emission
reported in DJ96 would be a direct proof for 
the synchrotron origin of the ``MeV bump''.

\section{Bremsstrahlung and $\pi^{0}$-decay Gamma Rays ?}

The second feature seen in Fig.~1 and Fig.~2 is the deficit in  the 
predicted IC fluxes compared to the reported fluxes  
at GeV energies\footnote{In Figs.~2,3 we show the  spectral points  reported
by Nolan et al. (1993) which are somewhat different from the 
results of the latest analysis (Fierro 1996)
shown in Fig.~1. The  
irregularities in the 1-10 GeV region could be due to the 
deficit of sufficient photon statistics in smaller energy bins 
used in the analysis of Fierro (1996).}.  
Indeed,  the expected IC $\gamma$-ray flux in this energy region 
is estimated   within  $20 \%$ accuracy  as
$I(1-10  \, {\rm GeV}) \simeq 10^{-8}
(\bar{B}/0.3 \, \rm mG)^{-1.3} \, \rm ph/cm^2 s$ (AA96).  
For the equipartition magnetic field $B \simeq 0.3 \, \rm mG$
this flux is by a factor of 5 below of the one measured by EGRET.
In order to explain the measured fluxes in this energy region,
an additional component of GeV radiation
was suggested in DJ96, which the 
authors in their best-fit ``synchrotron+IC''
model call as a {\it second IC power-law component}.
Meanwhile,  to  increase the IC flux one has to suppose 
for the magnetic field  
in the radio nebula (where the low energy $\gamma$-rays are produced)
$B \sim 10^{-4} \, \rm G$.  
However, this is a rather uncomfortable value since it would   
imply the energy in radio electrons 
$W_{\rm e} \simeq 7 \cdot 10^{48} \, \rm erg$, while 
the energy in the magnetic field were  by a factor of 30 smaller,
which makes rather problematic the confinement of electrons in the nebula. 
Thus, if the high EGRET flux above 1~GeV really originates in
the nebula,  then invocation of other mechanism(s) for additional  
production of $\gamma$-rays, e.g. due to 
interactions of relativistic particles 
with the ambient nebular gas (AA96), seems  unavoidable.

\vspace{2mm}
\noindent
{\bf Bremsstrahlung gamma-rays.}~~For the mean gas 
density in the nebula
$\bar{n} \approx 5 \, \rm cm^{-3}$, 
the flux of the bremsstrahlung $\gamma$-rays cannot exceed $15\, \%$ of
the flux of IC $\gamma$-rays.
In fact, in the Crab Nebula the gas is concentrated
mainly in  dense filaments  where
$n \sim 10^3 \, \rm cm^{-3}$ (Davidson and Fesen 1985). 
In the case of uniform distribution
of relativistic electrons throughout the nebula
the {\it effective} gas density
is defined by the mean density of the nebula,
$n_{\rm eff} \approx \bar{n}$.
However,
if electrons are trapped, at least partially, in the regions of
high density, i.e. if they propagate
inside the filaments {\it slower} than outside, 
then $n_{\rm eff} \gg \bar{n}$ (for details see AA96). 
The fluxes of bremsstrahlung $\gamma$-rays calculated for
$n_{\rm eff}=50 \, \rm cm^{-3}$
are shown in Fig.~3.  The  contribution of the 
``amplified'' bremsstrahlung flux   
not only could explain  the measured GeV $\gamma$-ray
fluxes, but also would  significantly  modify 
the spectrum at very high energies. Indeed, 
in the energy range between $100 \, \rm GeV$ and
$10 \, \rm TeV$ the superposition of the IC and 
the ``amplified'' bremsstrahlung components results in almost 
power-law spectrum with an index
$\alpha_{\gamma} \simeq (2.5-2.7)$ in contrast to 
the curved IC $\gamma$-ray spectrum  {\it alone},  which is  
hard at $E\simeq 100 \, \rm GeV$ ($\alpha_\gamma \simeq 2.0$),  but becomes
significantly steeper at higher energies 
($\alpha_\gamma \simeq 2.7$ at $E \simeq 10 \, \rm TeV$).

\begin{figure}[t]
\vspace{5.5 cm}
\includegraphics{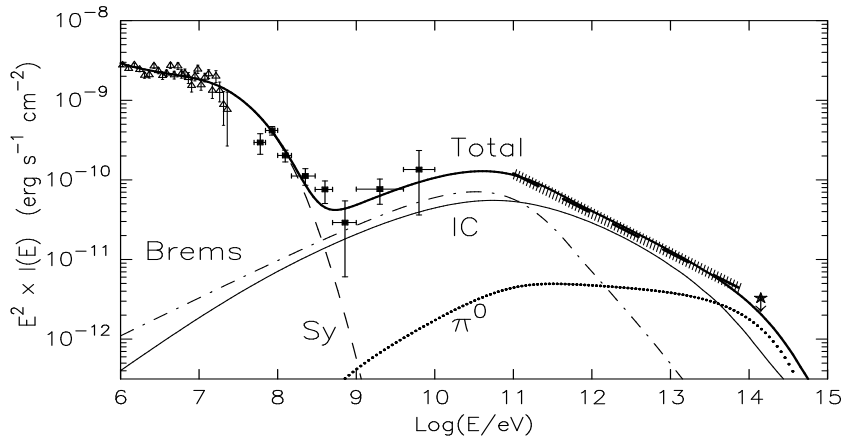}
\caption{\protect \footnotesize The contributions of different 
$\gamma$-ray production mechanisms to the total nonthermal radiation of the  
Crab Nebula. The Synchrotron and IC components are the same as in Fig.2.
The bremsstrahlung and $\pi^0$-decay $\gamma$-ray fluxes are calculated
for $n_{\rm eff}=50 \, \rm cm^{-3}$ (see text).}
\end{figure}

\vspace{2 mm}
\noindent
{\bf $\pi^{0}$-decay gamma-rays.}~
Interactions of the  nucleonic component of accelerated
particles (which may in principle acquire significant part of the power
of relativistic wind at the reverse shock; see Arons 1996),  
with the ambient gas lead to production of $\gamma$-rays through
secondary $\pi^0$-decays. However, since the average gas density
in the nebula is low,  the contribution of this mechanism 
to the $\gamma$-radiation could be detectable only in the case of  
partial  confinement of relativistic particles in the  filaments, 
so that $n_{\rm eff} \gg \bar{n}$. If so, the $\pi^0$-decay 
$\gamma$ rays may effectively show up (on top of the steep
IC spectrum)  at multi-TeV  energies (AA96).
From this point of view, the recently reported detection of 
up to 70~TeV $\gamma$-radiation from the Crab
by the CANGAROO collaboration (Tanimori et al. 1998) 
may have significant implications 
concerning the content of the wind and propagation/interaction
of accelerated particles in the filaments. 
Although the CANGAROO fluxes do not contradict, 
within the uncertainties in the nebular magnetic field, 
to the IC origin of the radiation, the 
reported differential power-law spectrum 
from several TeV to 50 TeV with an index 
$\alpha_\gamma = 2.53 \pm 0.18$
seems to be essentially harder than the 
predicted IC spectrum
($\alpha_\gamma \simeq 3$ at 30 TeV). 

In Fig.~3 we show the spectrum of $\pi^0$-decay $\gamma$-rays
calculated for the power-law differential spectrum of 
accelerated protons with $\alpha_{\rm p}=2.1$,  
exponential cutoff at  $E= 10^{15} \, \rm eV$ 
and significant flattening below 
$E \sim \, \rm 1~TeV$, 
as it is expected from the wind acceleration models.
For $n_{\rm eff}=50 \, \rm cm^{-3}$  used in Fig.~3, 
the shown $\pi^0$-decay fluxes correspond to total energy in accelerated 
protons 
$W_{\rm p}=1.5 \cdot 10^{48} \, \rm erg$, 
a quite acceptable amount from the point of view of 
the energy budget of the Crab. 
It is intersting to note that
for the chosen parameters  the superposition of 3 components of 
radiation -- ``IC+bremsstrahlung+$\pi^0$'' -- results in the
power-law spectrum with $\alpha_\gamma \simeq 2.5$ 
over the entire energy range from 100~GeV to 100~TeV. 
This spectrum significantly differs from the
pure IC spectrum, and provides a better fit to the reported data the 
compilation of which is presented in Fig.~2 and 3 by the hatched zone.
However, given the large uncertainties  in the reported $\gamma$-ray 
fluxes,  presently we could not  make a strong statement about the 
existence of the bremsstrahlung and $\pi^0$ signatures in the Crab spectrum.

\section{Discussion}

The Crab Nebula, one of the most comprehensively studied objects on the sky,
is detectable from 10 MHz radio waves to multi-TeV 
$\gamma$-rays. 
The multiwavelength observations of the Crab Nebula have already provided 
deep insight into the origin of the source dominated by nonthermal 
energy in the form of magnetic fields and relativistic electrons. 
Meanwhile many details remain still unresolved,
and should be addressed by future observations.    
Importantly, the fluxes of the source
exceed, by at least one order of magnitude, the sensitivities
of the current or planned instruments at practically
all frequences of the observed spectrum.
This ensures further significant progress in understanding
of complex processes of interaction of 
the relativistic pulsar wind with the nebula.   
Below we briefly outline some problems which could be answered,
hopefully in the near future,  by new  observations 
in different energy bands.

\vspace{2 mm}
\noindent
{\bf Probing the magnetic field and
electrons in the synchrotron nebula.}~~
The most informative frequency band
to probe the  acceleration site(s) and the character of
propagation of the {\it wind electrons}
in the nebula is the X-ray domain. The AXAF, with its subarcsecond
imaging capability and excellent spectral resolution
(Elvis 1997), seems to be an ideal instrument for such study.
Meanwhile the synchrotron data alone , strictly speaking, tell us only
about the product of the energy densities of the
magnetic field and relativistic  electrons. These parameters
could be disentangled using additional information
contained in the IC $\gamma$-rays produced by the same electrons.
Although the limited angular resolution of $\gamma$-ray detectors does not
allow, at least presently, mapping of the source on subarcmin scales,
the measurements of integral fluxes of IC $\gamma$-rays
at different energies, being coupled with synchrotron radiation in
relevant energy bands,
could compensate this disadvantage.

Indeed,  since
the TeV $\gamma$-rays are produced by IC scattering of
electrons responsible for  the observed keV X-rays,  the estimate
of the magnetic field based on keV/TeV data relates to the
central $r \sim 0.5 \, \rm pc$ region of the Crab Nebula.
The model-independent estimate of the magnetic field in the outer
parts of the (optical) nebula can be provided by
measurements of $\gamma$-ray fluxes at
$E \sim 100 \, \rm GeV$.
Similarly, the fluxes  of $E \geq 10 \, \rm TeV$ $\gamma$-rays,
combined with hard X-ray data ($E \geq 100 \, \rm keV$),
could allow determination of the  magnetic field 
in the vicinity of  the wind shock front at $r \sim 0.1 \, \rm pc$.
Remarkably, as far as the target photon fields for IC $\gamma$-rays
are well known, the accuracy of the estimate of the
magnetic field  could be better than $25 \%$, provided that the
$\gamma$-ray fluxes are measured with accuracy better than $50 \%$.
Despite the current large uncertainties
in the reported TeV  $\gamma$-ray fluxes, caused essentially
by the uncertainties in the absolute calibration
of the telescopes, it is believed that such accuracy
will be reached soon by imaging Cherenkov 
telescopes like Whipple, HEGRA, and
CAT \footnote{for description of
current and planned ground-based $\gamma$-ray detectors
mentioned in this paper see e.g. the recent reviews by 
Aharonian and Akerlof (1997) and Fegan(1997).}.

The detection of the Crab well beyond 10~TeV by the CANGAROO group
is very encouraging, but it would be very important 
to have independent measurements in 
this energy region.   
Also, it is expected that in 1 or 2 years the $\gamma$-ray flux 
measurements from the Crab will be extended down to 30~GeV by 
low-energy threshold atmospheric Cherenkov detectors  
like STACEE and CELEST. Besides of the determination of the magnetic
field in the optical nebula, good spectroscopy 
at these energies could provide important information about the 
spectrum of electrons in the transition region from the {\it radio}
electrons to {\it wind} electrons. 

While at energies between 1 and 10 TeV 
the IC scattering dominates over all other possible radiation mechanisms,  
at energies below 1~TeV and above 10~TeV 
other processes connected with interaction of the 
relativistic particles with the nebular gas might contribute to the production 
of $\gamma$-rays as much as the IC does.
Therefore determination of the magnetic field based on $\gamma$-ray
fluxes in these energy regions requires separation 
of the IC contribution from the possible  
contamination due to $\gamma$-rays of other origin.  
The shape of the spectrum of IC $\gamma$-rays 
is rather stable to the basic parameters of the nebula. 
While at  GeV energies the IC spectrum is very hard, 
with power-law index $\alpha_\gamma \approx 1.5$, in the VHE region
the spectrum gradually steepens from $\alpha_\gamma \approx 2$ at
$E \sim 100 \, \rm GeV$ to  $\alpha_\gamma \approx 2.5$ at
$E \sim 1 \, \rm TeV$, and $\alpha_\gamma \approx 2.7$ at 
$E \sim 10 \, \rm TeV$ (see Fig.~1). 
Confirmation of this spectral shape seems  one of the 
important issues to be addressed by  new observations,
in particular by the stereoscopic systems of imaging
atmospheric Cherenkov telescopes  
which provide good spectrometry with  
energy resolution $\leq 20 \%$, and high precision of 
localization of the point VHE sources 
with subarcmin accuracy (Aharonian and Akerlof 1997).
Although this still is not sufficient for adequate 
study of  the angular structure of the VHE $\gamma$-ray production region
in the Crab, in combination with expected large photon statistics it  
could give an answer whether the observed $\gamma$-rays are produced in
the {\it nebula}, or should they be attributed to
the unpulsed radiation of the pulsar (Cheung and Cheng 1994,
Bogovalov and Kotov 1995, Stepanian 1995).

\vspace{2 mm}
\noindent
{\bf Interaction of accelerated particles with filaments.}~~
Although the uncertainties in the reported fluxes at energies
1-10 GeV do not allow one to claim about the strong conflict
between the observations and the prediced IC $\gamma$-ray
fluxes, the confirmation of high EGRET fluxes by future
observations, e.g. by the GLAST (Bloom 1996), would require
more effective mechanism(s) responsible for
$\gamma$-ray production in this energy region.
The bremsstrahlung of {\it radio} electrons seems
to be an intriguing possibility. In particular, it
implies an effective confinement of the relativistic
particles in dense filaments to provide
sufficiently high  {\it effective} gas density
(``seen'' by relativistic particles) $n_{\rm eff} \geq 50 \, \rm cm^{-3}$.
In its turn,  this implies essentially different speed
and character of propagation of relativistic particles in the
filaments and  in the rest of the nebula.
                                                               
The enhanced rate of interactions of  
relativistic particles with the gas due to their
confinement in high density regions, opens  possibility
to probe the content of the pulsar wind by searching for $\pi^0$
$\gamma$-rays in the spectrum of the Crab at very high energies.
Indeed, for the {\it effective} gas density 
$n_{\rm eff} \geq 50  \rm cm^{-3}$,  the total energy of  accelerated
protons (nuclei) $W_{\rm p} \leq 2 \cdot 10^{48} \, \rm erg$,
quite reasonable amount for the energy budget of the Crab,  
would be sufficient for significant modification (hardening) 
of the resulting ``IC+$\pi^0$'' spectrum.  If confirmed by future
accurate spectroscopic measurements, the relatively hard spectrum 
of $\gamma$-rays above 10~TeV reported by CANGAROO
($\alpha_\gamma \leq 2.7$), 
will indicate the acceleration of nucleonic component
in the Crab up to energies $10^{15} \, \rm eV$, as well as 
high concentration of relativistic particles in dense gas regions.
Interestingly, since this will  unavoidably result in   
the enhanced bremsstrahlung $\gamma$-rays as well,
the total ``IC+bremsstrahlung+$\pi^0$''  
radiation spectrum is expected to be of almost 
single power-law form  above 100~GeV over three decades in energy.

\vspace{2 mm}
\noindent
{\bf Origin of hard (MeV) synchrotron radiation.}~~
The apparent steepening in the 
synchrotron spectrum above 100~keV  followed by hard MeV
radiation  (``MeV bump'') can be naturally interpreted as
superposition of two different radiation components, both, 
most probably, of the synchrotron origin. 
If the first component is attributed to the diffuse emission
of the synchrotron nebula, the second component could 
originate in one or few compact structures, e.g. wisps or knots,
where the strong magnetic field ($B \geq 10^{-3} \, \rm G$)
could create favorable conditions for both effective 
acceleration of highest energy electrons and production of 
synchrotron radiation up to  several 100 MeV.   
The limited angular resolution of
hard X-ray/soft $\gamma$-ray detectors does not allow direct 
identification of these compact structures. 
On the other hand, 
since the cooling time of electrons producing 
MeV synchrotron radiation does not exceed the light 
crossing time of these compact structures, 
the detailed study of the spectrum and flux variability 
of radiation in this energy domain by future 
INTEGRAL (Winkler 1996) and GLAST (Bloom 1996) missions, 
could compensate, to some extent, the lack of spatial information.

Meanwhile, direct information about the site(s) of the 
synchrotron ``MeV bump'' could be obtained in soft X-rays  
by AXAF. Indeed, even in the case of extremely hard 
spectrum (e.g. 
monoenergetic or Maxwellian type) of electrons
responsible for the ``MeV bump'', the energy flux
of this component at 1-10 keV range cannot be less than      
$10^{-11} \, \rm erg/cm^2 s$, so it 
should be easily detected by AXAF even if this flux is
due to superposition of large number of subarcsecond 
structures. However, detection of these structures in X-rays
will be not yet enough to identify  them  as the sites of ``MeV bump''.
The important criteria  for such identification 
is the expected very hard spectrum at keV energies, namely,  
essentially harder than 
the X-ray spectrum of the surrounding diffuse synchrotron nebula.

\footnotesize

\section{References}

\re
Aharonian, F.A. and Akerlof, C.W., 1997, Annual. Rev. Nucl. Part. Sci., 47, 273
\re
Aharonian, F.A. et al., 1997, A\&A, 327, L5
\re
Atoyan, A.M. and Aharonian, F.A., 1996, MNRAS, 278, 525
\re
Arons, J. , 1996, Space Sci. Rev., 75, 235
\re
Bartlett, L.M., 1994, Ph.D. thesis, University of Maryland
\re
Bogovalov, S.V. and Kotov, Yu.D., 1993, MNRAS, 262, 75
\re
Borione, A. et al., 1997, ApJ, 481, 313
\re
Bloom, E.D., 1996, Space Sci. Rev., 75, 109
\re
Cheung, W.M. and Cheng, K.S., 1994, ApJ (Suppl), 90, 827
\re
Davidson, K. and Fesen, R.A., 1985, Ann. Rev. Astron. Astrophys., 23, 119
\re 
de Jager, O.C. and Harding, A.K., 1992, ApJ, 396, 161
\re
de Jager, O.C. et al., 1996, ApJ, 457, 253
\re
Elvis, M., 1997, to appear in `The Hot Universe', IAU Symposium 188.
\re
Fegan, D.J., 1997, J.Phys. G: Nucl. Part. Phys., 23, 1013
\re
Fierro, J.M., 1996, Ph.D. thesis, Stanford University.
\re
Gould, R. J., 1965, Phys.Rev. Lett., 15, 577
\re
Hester, J.J. et al., 1995, ApJ, 448, 240
\re
Jung, G.V., 1989, ApJ, 338, 972
\re
Kennel, C.F. and Coroniti, F.V., 1984, ApJ, 283, 694
\re
Marsden, P.L. et al., 1984, ApJ, 278, L29
\re
Mohanty, G. et al.,  1998, Astroparticle Physics, in press
\re 
Nolan, P.L. et al., ApJ, 409, 710 
\re
Rees, M. J. and Gun, J.E., 1974, MNRAS, 167, 1
\re
Stepanian, A.A., 1995, Nucl. Phys. B, 39A, 207
\re
Tanimori, T. et al. 1998, ApJ, 429, L33
\re
van der Meulen, R.D. et al., 1998,  A\&A, in press 
\re
Weekes, T.C. et al., 1989, ApJ, 342, 379
\re
Winkler, C., 1996, A\&A (Suppl.), 120, 637

\end{document}